\documentclass[aps,prl,showpacs]{revtex4} 
\usepackage{amsmath} % Enhanced mathematics
\usepackage{amssymb} % Enhanced mathematics
\usepackage{graphicx} % Include figure files
\usepackage{graphics}
\usepackage{epsfig}
\usepackage{slashed}
\newcommand\nn{\nonumber}
\newcommand\ba{\begin{eqnarray}}
\newcommand\ea{\end{eqnarray}}
\newcommand\alb{\begin{align}}
\newcommand\ale{\end{align}}
\newcommand\be{\begin{equation}}
\newcommand\ee{\end{equation}}

\usepackage{amsmath} % Enhanced mathematics 
\usepackage{amssymb}
\begin{document}

		\title{Soft rescattering in the timelike proton form factor within 
a spacetime scheme}
%		\title{Evidence of a systematic modulation pattern and 
 % chances of a large imaginary part in the proton time-like form factor} 

%\title{Evidence for rescattering effects with large contributions to 
%the imaginary part of the proton timelike form factor}

		\author{Andrea Bianconi} 
    \affiliation{\it Dipartimento di
	Ingegneria dell\!~$^\prime$Informazione, Universit\`a degli Studi di Brescia
	and Istituto Nazionale di Fisica Nucleare, Gruppo Collegato di
	Brescia, I-25133, Brescia, Italy}

		\author{Egle~Tomasi-Gustafsson} 
		\affiliation{\it IRFU, CEA, Universit\'e Paris-Saclay, 91191 Gif-sur-Yvette Cedex, France}

		\date{\today}

		\begin{abstract}
The annihilation of a lepton pair into a proton-antiproton pair, and the 
reverse process, allow for a measurement of the timelike proton form 
factors. This work aims at studying the corrections due to direct 
soft interactions 
between the proton and the antiproton before their annihilation or after 
their creation. The analysis is carried on in the spacetime formalism 
used by us in a series of recent works on timelike FFs. 
In particular the effect of $\bar{p}p$ annihilation into 
many-meson states is considered. We expect it to determine similar 
real and imaginary parts for the form factor near the $\bar{p}p$ 
channel threshold. We also discuss the possibility that the timelike FF 
differ by a phase in the hadron pair 
production and annihilation. 
                \end{abstract}
\maketitle

%%%%%%%%%%%%%%%%%%%%%%%%%%%%%%%%%%%%%%%%%%%%%%%%%%%%%%%

%%%%%%%%%%%%%%%%%%%%%%%%%%%%%%%%%%%%%%%%%%%%%%%%%%%%%%%%%%%%%%%%%%%%%%%
%%%%%%%%%%%%%%%%%%%%%%%%%%%%%%%%%%%%%%%%%%%%%%%%%%%%%%%%%%%%%%%%%%%%%%
%%%%%%%%%%%%%%%%%%%%%%%%%%%%%%%%%%%%%%%%%%%%%%%%%%%%%%%%%%%%%%
\section{Introduction}
\begin{figure}
\begin{center}
\includegraphics[width=8.5cm]{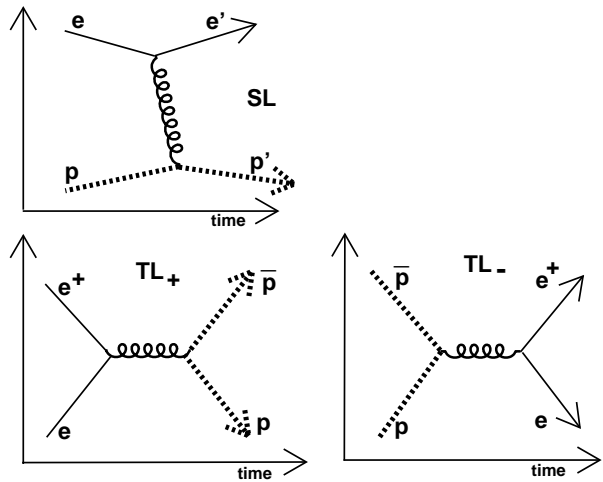}
\caption{
Reactions (\ref{eq:eq0}) (labeled as SL in the figure), 
(\ref{eq:eq0}) (TL$_+$), (\ref{eq:eq0})  (TL$_-$). In the one photon exchange approximation, electromagnetic 
FFs are functions characterizing the vertex coupling the virtual 
photon to the hadron current (thick grey dashed line in the figure). 
}
\label{Fig:reactions}
\end{center}
\end{figure}

The electromagnetic proton current is conveniently parametrized in terms of 
form factors (FFs).
FFs are experimentally extracted from elementary relations 
involving electrons and protons, assuming that the interaction occurs through 
the exchange of a virtual photon of four-momentum $q^2$.  The region of 
space-like (SL) momenta is investigated with  the elastic reactions:
\be
e^\pm + p \to e^\pm + p,
\label{eq:eq0a}
\ee
whereas the electromagnetic FFs of the proton in the time-like (TL) region 
(TLFFs) are accessible from the reactions 
\ba
&e^+ + e^- \to \bar p +p , 
\label{eq:eq1a}\\
&\bar p + p \to e^+ + e^- 
\label{eq:eq2a}
\ea
In the one-photon exchange approximation, these reactions are described 
by the diagrams of Fig. \ref{Fig:reactions}. They are completely 
defined by two FFs, $G_E$ and $G_M$. 
The unpolarized cross sections are  expressed in terms of the FFs squared 
in the SL region. In the TL region FFs are complex functions of $q^2$ 
and the annihilation cross section is expressed as function of FFs moduli 
squared \cite{Zichichi:1962ni},  see also 
\cite{Pacetti:2015iqa,Denig:2012by} for recent reviews. 

In the SL region, FFs have long been determined through the 
Rosenbluth method \cite{Rosenbluth:1950yq} i.e., the measurement 
of the unpolarized cross section at fixed $q^2$ for different angles. 
The availability of high duty cycle electron accelerator, highly polarized 
electron beams, as well as large solid angle spectrometers and 
development of polarimetry in the GeV region, recently allowed 
to apply the Akhiezer-Rekalo method:
\cite{Akhiezer:1968ek,Akhiezer:1974em}: the measurement of 
the recoil proton polarization in the scattering plane, from 
the elastic $\vec ep$ reaction, where the electrons are 
longitudinally polarized, allows to access directly the ratio $ G_E /G_M$, 
being more sensitive to a small electric contribution to the cross section.

%%%%%%%%%%%%%%%%%%%%%
\subsection{Effective proton timelike form factor - leading trend}
%%%%%%%%%%%%%%%%%%%%%%

The experimental and theoretical investigations of the TLFF 
are less advanced than for the spacelike (SL) case.  In 
particular, the experimental separation of the electric and 
the magnetic FF has not been possible in the TL region, because of 
the available limited luminosity. 
The precision on the cross section $\sigma$ of the reactions  
(\ref{eq:eq1a},\ref{eq:eq2a}) has only 
allowed for the extraction of the squared modulus of a single effective 
FF, $F_p$,  that for reaction \ref{eq:eq1a} is defined 
as \cite{Bardin:1994am}:
\be
|F_p|^2=\displaystyle\frac{3\beta q^2 \sigma}
{2\pi\alpha^2 \left(2+\displaystyle\frac{1}{\tau}\right)}, 
\label{eq:Fp}
\ee
where $\alpha=e^2/(4\pi)$, $\beta=\sqrt{1-1/\tau}$, 
$~\tau={q^2}/(4M^2)$, $q^2$ is the squared invariant mass of the colliding 
pair, and $M$ is the proton mass. 
For reaction (\ref{eq:eq2a}) we have a small difference 
in the definition of effective FF in terms of $\sigma$ 
(the final state phase space is not present anymore). 
The effect of the Coulomb singularity 
of both cross sections at the $\bar{p}p$ 
threshold is removed by the $\beta$ factor: 
$\beta$ $\rightarrow$ 0 for $q$ $\rightarrow$ $2M$, so that $\beta \sigma$ 
is finite and the effective FF is expected to be finite at the 
threshold. 
%%%%citazioni nella caption. 

\begin{figure}
\begin{center}
\includegraphics[width=8.5cm]{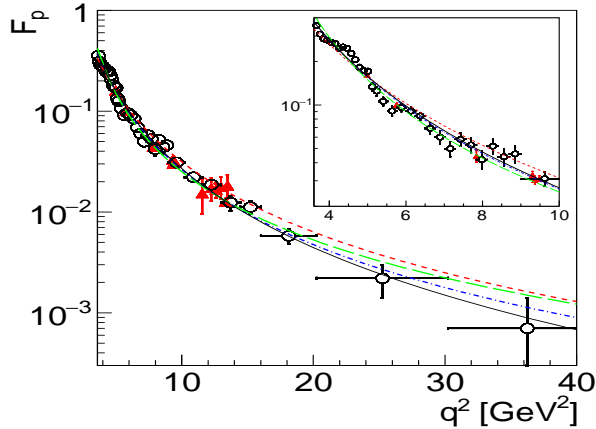}
\caption{Most recent data on the TL proton generalized FF as a function of $q^2$, from Ref. \protect\cite{Lees:2013xe,Lees:2013uta} (black open circles), 
Refs. \protect\cite{Ablikim:2005nn}  (red triangles),
together with the calculation from Eq. (\ref{eq:qcd}) (blue dash-dotted line, 
from \cite{Ambrogiani:1999bh,Lepage:1979za}), and from 
\cite{Shirkov:1997wi,Kuraev} (red dashed line), 
 \cite{Brodsky:2007hb} (green long-dashed line), and 
\cite{TomasiGustafsson:2001za} (black solid line). 
}
\label{Fig:WorldData}
\end{center}
\end{figure}

This effective TLFF has been measured by several experiments for $q^2$ 
ranging from the threshold $(2 M_N)^2$ to about 36 GeV$^2$. 
In Fig. \ref{Fig:WorldData} the most recent and precise results are 
reported, from the Babar  \protect\cite{Lees:2013xe,Lees:2013uta} and 
BES III collaborations  \protect\cite{Ablikim:2015vga}. These data have been fitted by some parameterizations.  
Four of them are reported in the figure. 
Details about these fits and the best fit values of their parameters 
can be found in our previous work \cite{Bianconi:2016owb}. The most visible  
feature is a strong power-law fall at increasing $q$. For example, 
in the experimental papers before 
the year 2006, the simple function \cite{Ambrogiani:1999bh,Lepage:1979za}: 
\be
|F_{scaling}(q^2)|=\displaystyle\frac{\cal A}{(q^2)^2\log^2(q^2/\Lambda^2)} 
\label{eq:qcd}
\ee  
was frequently used, suggesting a $1/q^4$ trend 
apart for logarithmic corrections. 
The TLFF data from the BABAR 
collaboration \cite{Lees:2013xe,Lees:2013uta} 
extending from the threshold to $q^2$ $\approx$ 36 GeV$^2$, look even 
steeper than this. 
%%%%%%%%%%%%%%%%%%%%%%%
\subsection{Modulation of the leading trend}
%%%%%%%%%%%%%%%%%%%%%%%%%%%%
For $q< $4 GeV the data also 
show oscillating 10 \% modulations around this regular trend. In our  
works \cite{Bianconi:2015owa,Bianconi:2016owb} 
we have 
fitted the BABAR data with  
\be
F(p)\ \equiv\ F_0(p)\ +\ F_{osc}(p), 
\label{eq:diff}
\ee
where $p =p(q)$ is the relative three-momentum of the final hadron pair, 
$F_0(p)$ is a standard fit expressed in terms of $p(q)$, and 
the modulation term $F_{osc}(p)$ is parameterized as
\be
F_{osc}(p)\ \equiv\ A\ e^{- Bp}\ \cos(C p\ +\ D). 
\label{eq:eqdif}
\ee
The precise values of the parameters depend on the 
fit that is chosen as leading term $F_0$ (see 
\cite{Bianconi:2016owb} for the four different choices). 
In all cases $A$ has magnitude $\simeq 0.1$, and $D$ $\approx$ 0. 
The smallness of $A$ suggests that the oscillations are due to some 
perturbation of a ``leading'' physics connected to the term $F_0$. 
$D$ $\approx$ 0 is an indication  
that the first oscillation is also a threshold enhancement, like 
those  found in $e^+e^-\rightarrow\bar{n}n$,  
$e^+e^-\rightarrow\bar{\Lambda}\Lambda$ and other 
production processes of neutral baryon 
pairs \cite{BESIII:2010ad,Pakhlova:2008vn,Ablikim:2006dw,Ablikim:2004dj,Bai:2003sw,Amsler:1994ah}.  

\subsection{Hard and soft physics}

At large $q^2$ things should simplify: the correction $F_{osc}$ should 
disappear, and $F_0$ should converge to the  
quark counting rules: TLFF $\propto$ $1/q^4$ (with logarithmic corrections), 
as for the case of the  
SLFF asymptotic \cite{Matveev:1973uz,Brodsky:1973kr}. 
%%% Tutta la parte sui modelli ``leading'' è stata riassunta in una lista 
%%% di citazioni senza commenti. 
However, soft processes are expected to heavily affect the 
finite-$q$ deviations from the $1/q^4$ rule and to determine both the 
FF magnitude and phase not predictable by  the quark counting rules. 
This has prompted several studies of the 
nonperturbative aspects of the TLFF and/or models for them and/or 
approaches to measurements 
\cite{Brodsky:2007hb,
Gousset:1994yh,
Bijker:2004yu, Adamuscin:2005aq,
Belushkin:2006qa,Lomon:2012pn,       
Bianconi:2006wg,Bianconi:2006bb, 
Gakh:2005wa,Gakh:2005hh,
deMelo:2003uk,deMelo:2005cy, 
Kuraev:2011vq}. 

These studies were targeted at the leading features of the 
data shown in Fig. \ref{Fig:WorldData}, the ``regular'' behavior 
reproduced by $F_0$. Concerning the 10\% oscillations of 
Eqs. (\ref{eq:diff}-\ref{eq:eqdif}), there are interpretations 
proposed by us 
\cite{Bianconi:2015owa,Bianconi:2016owb} and by \cite{PhysRevD.92.034018}, 
but this feature of the TLFF is still poorly explored. 

%Interference in spacetime and poles in $q$ 
%could be alternative ways to describe a similar mechanism. 
As observed, the first oscillation is a threshold enhancement. This 
phenomenon has been studied in the electroproduction 
of neutral hadron pairs, where it is more evident.  
In Ref. \cite{Haidenbauer2014102} an explanation in terms of nucleon-antinucleon 
strong-force potentials is proposed. Ref. \cite{Baldini:2007qg} focusses on local 
electric interactions between quarks and antiquarks of the two 
baryons. This is equivalent to a reciprocally induced  
electric polarization of the interacting spin-1/2 hadron and antihadron, 
a mechanism that has also been also used in Ref. \cite{Bianconi:2014era} 
to explain enhancements of the low-energy rates for processes 
involving antineutrons. Whether is electrostatic or strong, in all cases 
what is suggested is an interaction between hadrons (hadron-antihadron 
potentials) just before their annihilation or just after their creation. 
%%%%%%%%%%%%%%%%%%%%%%%%%
\subsection{Antiproton-nucleon interactions}
%%%%%%%%%%%%%%%%%%%%%%%%%%
Interactions between a nucleon and an antinucleon are by themselves an
important subject of nuclear physics. In particular data on the $\bar{p}$-nucleon 
and $\bar{p}-$nucleus annihilation process at low energies 
%dal PRC
(see \cite{Balestra:1985kn,Balestra:1989rb,Bizzarri:1973sp,Bruckner:1989ew,Balestra:1984wd,Balestra:1985wk,Bertin:1996kw,Benedettini:1997fk,Zenoni:1999su,Zenoni:1999st,Bianconi:2000nh,Bianconi:1999vq,Bianconi:2011zz,Aghai-Khozani:2018hnb})
%(H.Aghai-Khozani et al, Nucl.Phys.{\bf A 970}, 366-378 (2018) )
and on antiprotonic nuclei 
%nuovo
 \cite{Trzcinska:2000uu},    
%A.Trzcinska et al, Nucl.Phys. {\bf A 692} 176-181 (2001) ) 
as well as  related theoretical 
analyses 
%dal PRC
\cite{Bruckner:1989ew,Bianconi:2000ap,Batty:2000vr,Friedman:2014vva,Lee:2015hma,Wycech:2007jb,Uzikov:2011tz}
%(T.-G. Lee, C.-Y. Wong, Phys. Rev. {\bf C 93}, 014616 (2016)..  
%S. Wycech, F.J. Hartmann, J. Jastrzebski, B. Klos, A. Trzcinska, and T. von Egidy, Phys.Rev.{\bf C76}, 034316 (2007).. 
%Yu. N. Uzikov, J. Haidenbauer, and B. A. Prmantayeva Phys. Rev. {\bf C 84}, 054011 (2011). )
 show that:

(a) Annihilation cross sections are large taking the unitarity limit 
as a reference. 
This means that in some partial waves a relevant part 
of the incoming projectile flux disappears in the interaction process: 
the proton is a black-sphere absorber for an antiproton flux, 
at any energy. 
Elastic scattering is present, mainly 
of diffractive origin (for momenta over 100-200 ~MeV/c), and of refractive 
repulsive hard-core nature near threshold or just below it (antiprotonic 
atoms). This repulsive character is an indirect consequence of the large absorption, 
and not of a repulsive potential: the need of a regular behavior of the 
wave function at the border of the region where it is suppressed by 
absorption determines a strong reflected component. 

(b) A proton 
and an antinucleon do not overlap. When their surfaces come in touch a 
quark from the proton and an antiquark from the antiproton easily 
rearrange into a meson leaving an unstable system followed by annihilation 
into multi-meson states. This clearly hinders the chances 
that three quarks and three antiquarks may gather in a region of size $1/q$ 
(0.1 fm at threshold), that is a precondition for converting all of them  
into a single lepton pair. The same problem is present in the 
reverse process of $\bar{p}p$ production, since the two hadrons should 
be created in a configuration where their chance to exist is very 
small. 

(c) The available data on low-energy $\bar{p}p$ processes 
are normally described in 
terms of optical potential analyses within a non-relativistic formalism.  

%%%%%%%%%%%%%%%%%%%%%
\subsection{Aim of the present work}
%%%%%%%%%%%%%%%%%%%%%%

In a previous publication \cite{Bianconi:2016bss} we 
discussed the physics of the proton TLFF in spacetime. 
In the SL case the FFs
in the Breit frame ($q_0=$ 0, no energy transfer) are interpreted as 
Fourier space transforms of stationary charge and current distributions. 
We studied a corresponding 
interpretation for the timelike FF in the CM frame 
of the $e^+e^-$ collision where 
the photon has zero three-momentum (infinite space wavelength). What one 
finds is a distribution in time of the photon-quark-antiquark 
creation vertexes. Such distribution and the space charge distribution 
are the projections respectively on the time axis and on the three-space of one 
and the same four-dimensional spacetime distribution $F(x)$. 
 
For simplicity, we avoided the rescattering problem, since 
this is a broad subject and deserves a dedicated work. This is  the 
subject of the present paper. We use the word 
``rescattering'' to describe interactions between the hadrons 
both in reaction \ref{eq:eq1a} and  \ref{eq:eq2a} where, strictly speaking, 
it should be defined as ``pre-scattering''. 

Since it may be ambiguous to decide which processes have to be classified 
as rescattering, a specific definition 
will be given in the next section. Let us anticipate 
that a special stress will be given to elastic rescattering, to charge exchange 
and to annihilation. 

%Clearly this photon cannot test any charge density like in the SL case, 
%since a time-like momentum 
%can test time distributions of events, but not space 
%distributions. So any effect related to space separation 
%of opposite-sign electric charges is not detectable by it. 

%In a quantum field theory, however, the electric 
%charge is the strength of the coupling between a photon and a fermion 
%current, so we may think at $R(t)$ as an amplitude for 
%creating (or annihilating) elementary charge-anticharge pairs 
%at the time $t$. So, ``time charge  
%distribution'' can be understood as ``distribution in time of vertexes 
%$\gamma^* \rightarrow charge + anticharge$''. 

%We analyzed this concept both in general and within a quark scheme 
%(like in fig.\ref{Fig:Feynmann}), 
%found relations between the time and space ``charge'' densities 
%$R(t)$ and  $\rho(r)$,  and gave 
%some examples for $R(t)$ inspired by the phenomenology (power 
%laws, monopole and dipole, asymptotic behavior with counting rules). 

%%%%%%%%%%%%%%%%%%
\section{Soft rescattering}
%%%%%%%%%%%%%%%%%%%%%%
%A key point is to distinguish which processes ahould 
%belong to the ``rescattering'' class, and which ones should not. 

In a previous work \cite{Bianconi:2016owb} we introduced the role of rescattering in TLFFs, where an optical 
potential model was used to explain the oscillations of the effective proton TLFF above threshold. 
A basic assumption of that work 
was that rescattering was a small perturbation, while the most evident behavior 
of the FF (a very fast but regular decrease at increasing $q$) was 
due to a ``leading'' mechanism. The purely phenomenological background 
for this was the fact that the oscillations did not exceed 10 \% of 
the FF absolute value. 
Since the same ``smallness'' assumption will accompany all of the present work, we better need 
to define what we mean by ``small'', by ``leading'', and by ``rescattering''. 

There is no easy way here to split the 
interaction Hamitonian into a form like $V$ $=$ $V_F$ $+$ $V_R$ 
where separate terms are responsible for hadron formation and 
rescattering between physical hadrons. 
The only exception is when ``rescattering'' means Coulomb interaction 
between the final or initial hadrons, but we are interested in a broader 
class of phenomena, involving strong interactions. 

\begin{figure}
\begin{center}
\includegraphics[width=8.5cm]{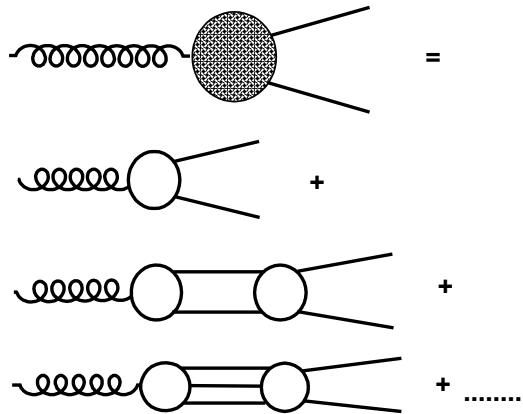}
\caption{
Amplitude for the formation of a $\bar{p}p$ pair as a sum of a ``bare'' 
amplitude plus rescattering corrections. These are of two kinds: elastic 
rescattering (that may be iterated 1, 2, ... times) and inelastic 
rescattering, with several possible intermediate states. In the present work 
we assume that the ``bare'' amplitude is actually a partially resummed 
amplitude calculated within some model, including all those rescattering 
processes that may be considered as hard, and all those soft processes 
that take place within the 3$q$ color singlet or within the 
3$\bar{q}$ singlet and do not involve singlet-singlet interactions.  
We consider as ``rescattering'' only those soft singlet-singlet 
interactions involving near-on-shell hadrons. 
}
\label{Fig:Sum}
\end{center}
\end{figure}

Within a perturbative and convergent scheme, the complete amplitude for a 
process like $e^+e^-$ $\rightarrow$ $\bar{p}p$ is a 
sum of a ``bare'' formation process plus an infinite set 
of amplitudes including rescattering events, as described 
in Fig. \ref{Fig:Sum}. This 
derives from the Born series expansion of the exponential that 
formally expresses the effect of the interaction potential: 
$exp(i\int V dt)$ $=$ $1$ $+$ $T \int iV dt'$ $+$ $T \int (iV)^2/2 dt'dt''$ 
$+$ $...$. So, when a perturbative scheme is available rescattering just means 
a class of higher order processes. 

Here a rapidly converging perturbative scheme is not available. 
Although much of the involved interactions can 
be treated within theories like PQCD or VDM, hard and soft strong interactions 
cannot be framed within one and the same perturbative scheme, and both are 
heavily involved in the pair formation process. 

To simplify our problem, and taking into account that our research line 
focusses on subleading behaviors, in the present work ``rescattering'' 
only means ``soft rescattering between physical hadrons''. 

To distinguish hard from soft processes, 
the point is not the nature of the involved particles (quarks vs hadrons, 
for example) but their offshellness. 
Even a model based on hadrons describes hard physics if the 
offshellness of these hadrons is hard-scaled. So, processes involving 
far-off-shell propagators are excluded from what we mean by rescattering. 

The ``bare formation'' of hadrons implies both hard and soft 
processes, as evident in the very basic and well known 
scheme where the reaction $e^+e^-\to \bar{p} p$ requires first 
the formation of three $\bar{q}q$ pairs within a $1/q$-sized region via 
perturbative quantum chromodynamics (PQCD) hard processes, 
but then each subgroup of three quarks and three antiquarks 
must evolve toward a physical hadron configuration with a 
much larger radius, and this evolution is non-perturbative and soft. 
In this path we have an ``unavoidable'' rescattering: after the 
first one, two more $\bar{q}q$ pairs are formed, and this cannot take place 
without involving both forming hadrons. Here to speak of 
rescattering as something 
distinguished from the formation process is nonsense. 
So, everything that takes place in this stage is not rescattering. 

Starting from the time when two separate color singlets may be 
identified, further interactions between the two become ``avoidable''. 
That is, they will take place with some frequency, but a proton and 
an antiproton can be formed in their absence too. 
If in some ``avoidable'' rescattering the involved 
propagators are highly virtual, 
the two steps of $\bar{q}q$ formation and rescattering are 
within $1/q$ in spacetime and even in this case difficult to 
conceptually disentangle. 

Those ``avoidable'' 
processes that involve  only soft propagators take place on a larger spacetime 
scale and in this approximate sense they may be distinguished from the 
``formation'' processes. Because of the longer involved wavelengths, 
these processes involve hadrons rather than individual quarks. 

The last relevant point is that soft steps are also heavily involved 
in the formation processes of the individual hadrons. 
We assume that on a time scale $\gg 1/q$ it is possible : 

a) to identify two color singlets that, although are not yet 
a proton and an antiproton, will separately evolve into them; 

b) to distinguish the intra-singlet from the inter-singlet processes.  

Summarizing, those ``avoidable'' processes that are soft and 
may be classified as interaction between singlets are what we mean 
with ``soft rescattering''. 

This is illustrated in Fig. \ref{Fig:Sum}, where the full amplitued is 
shown as the sum of a ``bare formation'' amplitude and of processes 
where the bare formation (the first blob on the left) is followed 
by further interaction. The legs on the right of the bare formation 
blob are near-on-shell hadrons. Those rescattering processes that 
involve hadrons or quarks with virtuality $\sim$ $q$ are considered 
as a renormalization of the bare formation amplitude, and not 
rescattering, and so, in Fig. \ref{Fig:Sum} they are hidden inside the 
formation blob. 

So, for example a VDM ``bare'' process like 
$e^+e^- \rightarrow M^* \rightarrow \bar{p}p$, 
where $M^*$ is a vector meson that is highly offshell in the physical 
$\bar{p}p$ channel, has 
rescattering corrections like 
$e^+e^- \rightarrow M^* \rightarrow \bar{p}p \rightarrow M^* \rightarrow \bar{p}p$.  
Such a correction is not considered as ``rescattering'' in this work, since it involves 
a meson with a hard-scale virtuality. We consider the terms like this as  
a renormalization of the hard part of the TLFF. 
On the other side, in a process like 
$e^+e^- \rightarrow M^* \rightarrow \bar{n}n\rightarrow \bar{p}p$ the final state 
charge exchange $\bar{n}n \rightarrow \bar{p}p$  
is definitely 
``soft rescattering'' in our interpretation if it is due to 
the $t-$channel exchange of a pion. The same is true for diffractive 
scattering mediated by a $t-$channel pomeron. 

In particular, we consider 
potential scattering as soft, even in the special case of an 
optical potential. An optical potential averages $\bar{p}p$ 
interactions 
that can be 
highly inelastic into an energy-conserving interaction that 
produces regular changes of the $\bar{p}p$ wave function (neglecting the 
effects on the other channels, and averaging irregularities 
over a finite energy range). So it describes 
soft effects of processes that can be hard. 

\begin{figure}
\begin{center}
\includegraphics[width=8.5cm]{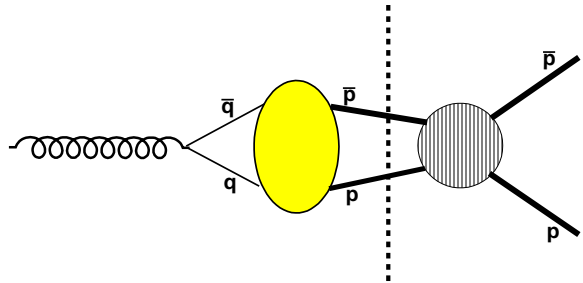}
\caption{
Process with initial formation of a quark-antiquark pair, then converted into 
a hadron pair according to a ``bare'' FF (left blob). The hadron 
pair interacts again (right blob, that may be t-channel or s-channel 
rescattering), and this adds a subleading correction to the FF. 
A cut on a physical state is possible in the intermediate 
di-hadron state, leading to an imaginary part of the FF. 
If the left blob contains physical intermediate states, then 
it may be crossed by another cut, and this will produce an imaginary 
part in the ``bare'' FF. If the rescattering 
blob is of s-channel kind, e.g. with formation of a single meson, 
the rescattering blob may be cut as well. So, several cuts are 
possible, but this work focusses on the cuts like the one showed here.  
}
\label{Fig:Cutkowsky}
\end{center}
\end{figure}

Rescattering allows for the presence of a Cutkowsky cut in the diagram for the 
photon-hadron vertex (see Fig. \ref{Fig:Cutkowsky}). This cut implies an 
imaginary part in the FF if it intercepts a state that may be 
physical at the considered kinematics. As evident from 
Fig. \ref{Fig:Cutkowsky}, such a cut may be present on the propagators 
of the hadrons in the intermediate state, but it may also cut the bare 
formation blob, since this hides hard rescattering processes. 
So an imaginary part may be  already 
present in the ``bare'' TLFF, but it surely receives contributions 
from hadronic states in presence of soft rescattering. In particular one may 
expect that an optical potential interaction contributes to the 
imaginary part of a FF. This point will be discussed at the 
end of this work. 

%If we consider the reverse reaction $\bar{p}p$ $\rightarrow$ $e^+e^-$, 
%initial state interactions play a heavy role in lowering the rate 
%of this reaction. Indeed, it has a strong competition from annihilation 
%reactions like $\bar{p}p$ $\rightarrow$ $mesons$. Near threshold 
%these take place when the two proton and antiproton 
%surfaces come close, without 
%the need of a true overlap between the hadrons. A quark and an antiquark 
%rearrange into a a meson and escape, leaving the remaining system unstable. 
%Reactions like this decrease much the chances for the proton and the 
%antiproton to reach some overlap, that is necessary to cluster all the 
%quarks and the antiquarks in a region of size $1/q$. 

\section{Four-dimensional Fourier transform of the FF}

The key tool for the investigation of FFs  is the four-dimensional Fourier 
transform \cite{Kuraev:2011vq}
\begin{align}
F(q)\ =\ \int d^4 x e^{iqx} F(x). 
\label{eq:FT_start}
\end{align}

Although $F(q)$ may be analyzed as a function of 
$q^2$ on all the complex plane, here we consider it as a function of 
all the 4 components of $q_\mu$, since the Fourier transform is 
separately performed with respect to each of them. So, in the following 
$x$ and $q$ mean $x_\mu$ and $q_\mu$. 
From a physical point of view, $F(x)$ is the amplitude $F(q)$ 
in another representation, that is the 
4-position of the photon-current vertex instead of the photon 4-momentum. 

\begin{figure}
\begin{center}
\includegraphics[width=9cm]{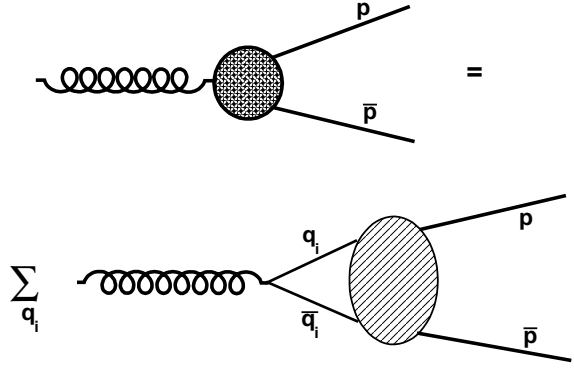}
\caption{
Unresolved and resolved level of analysis (see text), 
within a quark model. The TLFF in 4-coordinate representation $F(x)$ 
is a sum of amplitudes for 
converting a quark-antiquark pair that has been 
created in a 4-point into a proton-antiproton pair created 
in another 4-point, where $x$ $\equiv$ $x_\mu$ is the 4-distance of 
the two points. For the opposite process ($\bar{p}p$ annihilation 
into a virtual photon) it 
is an amplitude for finding a $\bar{q}q$ pair in a 4-point 
given a $\bar{p}p$ pair in another 4-point. 
} 
\label{Fig:FFquark}
\end{center}
\end{figure}

Although FFs are normally introduced as pieces of an amplitude, 
they may be considered amplitudes themselves. We need to distinguish 
between ``resolvable'' and ``unresolvable'' 
particles, as in Fig. \ref{Fig:FFquark}. 
A resolvable particle participates to a process with its internal 
structure, while an unresolvable particle is treated as elementary. 
As an unresolvable particle, the photon-hadron current interaction 
takes place in a single vertex $X_\mu$. At a resolvable level the 
photon-hadron interaction involves several variables $X_1^\mu,X_2^\mu,...$ 
associated to the internal hadron constituents (we omit the tensor 
indexes and just write $X$, $X_1$ etc).  
In particular let $X_1$ be the 4-position where the photon couples with 
an elementary (quark) current.

In coordinate representation, a FF $F(x)$ 
is the amplitude for the photon coupling taking place in a given 4-point 
$X_1$ at resolvable level, if the photon coupling at unresolvable level 
takes place in another point $X$, with $x=X_1-X$. Then of 
course things are made more complicate by the presence of more form 
factors entering a process, and by the fact that more $\bar{q}q$ 
pairs may couple with the photon. But this does not change the  
interpretation of a FF as an amplitude for seeing a hadron current 
at a deeper resolution level. 

In the SL case, and in the Breit frame where $q^\mu=(0,\vec q)$, 
the known non-relativistic interpretation of a form 
factor holds: 
\begin{align}
&F_{SL,Breit}(q)\ =\ \int d^3 \vec x\ exp(- i \vec q \cdot \vec x) 
\int dt F(t,\vec x) 
\ \equiv\ \int d^3 \vec x\ exp(- i \vec q \cdot \vec x) 
\rho(|\vec x|)
\label{eq:FSLBR}
\\
&\rho(|\vec x|)\ =\ \int dt F(t,\vec x). 
\label{eq:rho}
\end{align}
where 
$\rho(|\vec x|)$ may be read as a static charge density. Here it appears 
as a time 
average over the Fourier transform $F(x)$  $=$ $F(t,\vec x)$. 

In the TL case, and in the center of mass frame ($\vec q$ $=$ 0) 
we have a corresponding interpretation: 
\begin{align}
&F_{TL,CM}(q)\ =\ \int dt\ exp(i q t) \int d^3 \vec x F(t,\vec x) 
\ \equiv\ \int dt\ exp(i q t) R(t),
\label{eq:FTLCM}
\\
&R(t)\ =\ \int d^3 \vec x F(t,\vec x).
\label{eq:Rt}
\end{align}

It must be observed that in neither case one may access $F(x)$. Either 
one explores $\rho(\vec x)$ or $R(t)$, that are integrals of $F(x)$. So 
measuring TL and SL FFs leads to complementary pieces of 
information. 
%%%%%%%%%%%%%%%%%%%%
\section{The no-rescattering case.}
\begin{figure}
\begin{center}
\includegraphics[width=8.5cm]{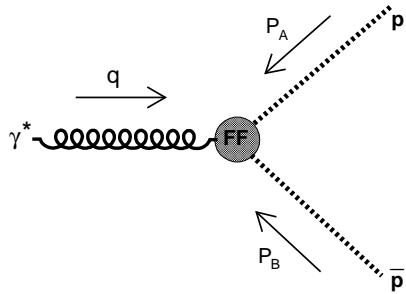}
\caption{
Diagram describing the three reactions (\ref{eq:table1}, 
\ref{eq:table2}, \ref{eq:table3}). Since these are actually 
physical channels of the same reaction, all of them may be described 
by the same diagram and by the same amplitude  
by changing the values of the components of the 
four-momenta $P_A$, $P_B$, $q$ and exploiting crossing symmetry. 
Formally, we consider these momenta 
as all entering. So, $q$ coincides with the physical four-momentum of 
the virtual 
photon in the channel $TL_+$ where a $\bar{p}p$ pair is created,  
while $q$ $=$ $-q'$, where $q'$ is the physical 
four-momentum of the virtual photon, in the reverse channel $TL_-$ 
where a virtual photon is produced 
by $\bar{p}p$ annihilation. Similar considerations apply to 
$P_A$ and $P_B$, see Eqs. (\ref{eq:table1},\ref{eq:table3}), for all 
the correspondences between formal arguments of the amplitude and 
physical momenta). 
}
\label{Fig:diagram}
\end{center}
\end{figure}

%%%%%%%%%%%%%%%%%%%%
The reactions from which the TL and the SL FFs are extracted 
are related by crossing symmetry (see Fig. \ref{Fig:reactions}). 
The one-photon exchange mechanism is assumed, so in the following ``FF'' 
stays for a renormalizing factor for the hadron-virtual photon vertex, as in 
Fig. \ref{Fig:diagram}. 
Factorizing out the lepton part of the process and the virtual 
photon propagation, we only consider the 
three-leg amplitude $A(q,P_A,P_B)$ 
describing the sub-processes 
\ba
&SL:&\ \gamma^*(q_\mu)\ +\ p(p_\mu) \rightarrow p(p_\mu'),
\label{eq:eq0}\\
&TL_+:&\ \gamma^*(q_\mu)\ \rightarrow p(p_\mu')\ +\ \bar{p}(\bar{p_\mu}'),
\label{eq:eq1}\\
&TL_-:&\ p(p_\mu)\ +\ \bar{p}(\bar{p}_\mu)\ \rightarrow \gamma^*(q_\mu'). 
\label{eq:eq2}
\ea
The four-momenta $q^\mu$, $P_A^\mu$, $P_B^\mu$ appearing as $formal$ 
arguments of $A(q,P_A,P_B)$ 
are all incoming as in Fig. \ref{Fig:diagram} 
so that the different reactions are distinguished by the 
expression of $q$, $P_A$ and $P_B$ in terms of the $physical$ 
momenta $q$, $q'$, $p$, $p'$, $\bar{p}$, $\bar{p}'$ 
(that have a positive time component if they are timelike):  
\begin{align}
&\gamma^* + p \rightarrow p'\hspace{1truecm}  
(SL:\ |q_0|\ <\ |\vec q|)\ 
&P_A\ =\ p,\ P_B\ =\ -p', 
\label{eq:table1} 
\\ 
&\gamma^* \rightarrow \bar{p} + p\hspace{1truecm}  
(TL_+:\ |q_0|\ >\ |\vec q|,\ q_0\ >\ 0)\  
&P_A\ =\ -p',\ P_B\ =\ -\bar{p}'.
\label{eq:table2} 
\\ 
&\bar{p} + p \rightarrow \gamma^*\hspace{1truecm}    
(TL_-:\ |q_0|\ >\ |\vec q|,\ q_0\ <\ 0)\  
&P_A\ =\ p,\ P_B\ =\ \bar{p},\ q\ =\ -q'.
\label{eq:table3} 
\end{align}
where 
in the TL region two reciprocally inverse reactions are possible, 
corresponding to $\bar{p}p$ annihilation into or creation from 
a lepton-antilepton pair. 

It is important not to confuse the four-momenta $P_A$ and $P_B$ as formal arguments  
of $A$, with their physical values $\pm p$, $\pm\bar{p}$...: 
the analytical continuation of $A(q,P_A,P_B)$ requires that 
this amplitude is described in terms of the same arguments in all the 
reaction channels and in the nonphysical regions  
(so, $q_0$ $<$ 0 in one of the two annihilation channels, and it is 
a complex variable in general). 
Being $A(q,P_A,P_B)$ invariant, it actually depends on $q$, $P_A$ and $P_B$ 
via their invariant products only, so these three four-vectors contain 
redundant information. However, in the following we keep the formal 
dependence of $A$ on them. 

Assuming a muon as a template for an unresolvable proton, $i.e.,$ 
a ``pointlike'' proton, 
the vertex matrix element for 
$\gamma(q)$ $+$ $\mu(p)$ $\rightarrow$ 
$\mu(p')$ is (using $\bar{u}\gamma^0$ $=$ 
$u^+$) 
\begin{align}
&A_{point\ SL}(q,p,p')\ =\ 
<\mu'| A_\nu(X) J^\nu(X)\ |\mu>\  =\ 
e \int d^4X\ 
e^{iqX}\ e^{-ip'X}\ e^{ipX}\ e_\nu\ \bar{u}(p') \gamma^\nu u(p)\ =\ 
\\
&=\ e \int d^4 X 
e^{iqX} e^{-ip'X} e^{ipX} 
\Big(
e_0\ u^+(p') u(p)\ -\ 
\vec{e}\ \bar{u}(p') \vec{\gamma} u(p)\Big) \ =\\ 
&=\   
\delta^4(q+p-p')\ \Big( T_{point\ charge}(q,p,p')\ -\ 
T_{point\ current}(q,p,p') \Big).  
\label{eq:a11_5}
\end{align} 
Exploiting that the amplitudes of the processes 
(\ref{eq:table2}) and (\ref{eq:table3}) 
are analytical continuations of the amplitude of (\ref{eq:table1}), 
we can write Eq. (\ref{eq:a11_5}) in a compact form describing all 
these processes: 
\begin{align}
A_{point}(q,P_A,P_B)\ \equiv\ 
\delta^4(q+ P_A+ P_B)\ \Big( T_{point\ charge}(q,P_A,P_B)\ -\ 
T_{point\ current}(q,P_A,P_B) \Big).  
\label{eq:a11_5_gen}
\end{align} 
where 
assigning to $q$, $P_A$, $P_B$ the values listed in 
Eqs. (\ref{eq:table1},\ref{eq:table2},\ref{eq:table3}),  
we obtain the amplitudes for the corresponding reactions. 
%$\gamma^*+\mu^+$ $\rightarrow$ $\mu^+$, 
%$\gamma^*$ $\rightarrow$ $\mu^+\mu^-$, 
%$\mu^+\mu^-$ $\rightarrow$ $\gamma^*$. 

Form 
factors may be introduced as scalar functions that multiply the previous terms, 
or linear combinations of these terms:  
\begin{align}
A(q,P_A,P_B)\ \equiv\ A_{charge}(q,P_A,P_B)\ -\ A_{current}(q,P_A,P_B)
\label{eq:a_split}
\\ 
\equiv\ \delta^4(q+P_A+P_B)\ 
\bigg(T_{point\ charge}(q,P_A,P_B)\ F(q)\ 
-\ T_{point\ current}(q,P_A,P_B)\ G(q) \bigg)
\label{eq:Amp_start}
\end{align}
where now this amplitude describes processes involving proton and antiproton 
instead of muons. The scalar FFs 
$F(q)$ and $G(q)$ depend on $q_\mu$ via the scalar $q^2$ $\equiv$ 
$q_\mu q^\mu$ only. Alternatively, one may rewrite the hadron four-current in 
the Gordon form, insert $F_1$ and $F_2$ and next combine them into 
$G_E$ and $G_M$, but the adopted procedure 
is simpler since it immediately highlights the term that is proportional 
to the charge density operator $u^+(p')u(p)$, and we will not work on the 
other component in the following.

Our further analysis \underline{only considers the FF $F(q)$ 
associated with the charge term}. 
So our starting equation is:
\begin{align}
A_{charge}(q,P_A,P_B) 
\equiv\ \delta^4(q+P_A+P_B)\ 
T_{point\ charge}(q,P_A,P_B)\ F(q) 
\label{eq:Amp_charge_start}
\end{align}
%%%%%%%%%%%%%%%%%%%%%%%%%%%%%%%%
\section{Translation invariance and Inner degrees of freedom.}
%%%%%%%%%%%%%%%%%%%%%%%%%%%%%%%%

We consider here the FF in a quark model framework, where it is 
assumed that the process of $\bar{p}p$ creation or annihilation is built  
around a quark-antiquark creation by the photon, or quark-antiquark 
annihilation into a photon, and several additional degrees of freedom  
enter the process, for example spectator quark coordinates. 

The amplitude describing how a $free$ (anti)proton with momentum $p$  
splits into a Fock state of $N$ constituents is 
\begin{align} 
\psi(X_1,X_2,....X_N)\ \equiv\ e^{ipX}\Phi(x_1,x_2,...x_N)
\label{eq:separation}
\end{align}
where the four-vector $X_i$ is the spacetime position of the i-th constituent, 
$X$ is a linear combination of all the $X_i$, expressing the 
spacetime position of the proton as a whole (the unresolved proton) 
and the four-coordinates $x_i$ are internal four-coordinates relative to $X$:  
\begin{align}
X\ =\ \sum w_i X_i,\ i=1,....N. 
\label{eq:coord1}
\\
x_i\ \equiv\ X_i-X.
\label{eq:coord2}
\\ 
\sum\ w_i x_i\ =\ 0, 
\label{eq:coord3}
\end{align}
where $w_i$ are weights that depend on dynamics (for example, on 
the longitudinal fractions or on the mass) within a given model. 

A key point is that because of spacetime translation invariance, 
a separation between relative and absolute coordinates leading to the 
factorization (\ref{eq:separation}) must exist, at the condition that the 
proton as a whole does not interact with the surrounding environment. 
This prerequisite is not present anymore if we introduce rescattering, 
that in our case means interactions 
between the proton and the antiproton before their annihilation or 
after their creation. Of course, an isolated set of particles may  
always be identified, unless very long range forces are included. This 
will lead to the separation of a degree of freedom that enjoys 
translational invariance. 

\begin{figure}
\begin{center}
\includegraphics[width=9cm]{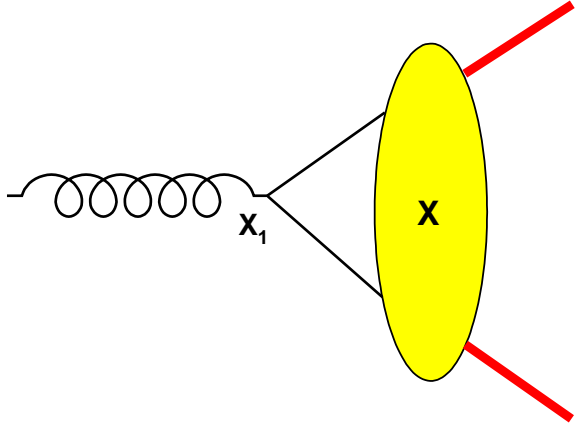}
\caption{
Diagram corresponding to Eq. (\ref{eq:invariance2}). The four-coodinate 
where the 
hadron pair is formed/annihilated (timelike process) or where the hadron 
is scattered (spacelike process) is $X$. At resolved level, the 
coordinate where the $\bar{q}q$ pair is formed/annihilated (TL), or where 
a quark is scattered (SL) is $X_1$. The electromagnetic FF 
quantifies the effect of the photon probe on the degree of freedom 
$x_1$ $\equiv$ $X_1-X$. 
} 
\label{Fig:Feynmann}
\end{center}
\end{figure}

We now consider first a diagram contributing to the FF in 
absence of rescattering, as in Fig. \ref{Fig:Feynmann}. 
So, $\Phi$ is a 
splitting amplitude for a free proton, and is 
a fully relativistic amplitude, 
where each four-coordinate has an independent time dependence. 

%In eq.\ref{eq:separation} $X$ is not the hadron center of mass in 
%nonrelativistic sense, since in eq.\ref{eq:coord1} 
%the positions of the partons are  
%taken at different times. But, 
%if the hadron current is not interacting with the environment,  
%a four-coordinate $X$ must exist that makes the factorization of 
%eq.\ref{eq:separation} possible, because the $exp(ipX)$ term 
%expresses the spacetime translation invariance of the 
%(anti)proton as a whole. 
%So, $X$ is the four-coordinate 
%of the unresolved proton and antiproton in the $\gamma^*\bar{p}p$ 
%vertex. 

Let us assume that the vitual photon creates a $\bar{p}p$ pair in 
a non interacting state. So, this state is described by the product 
$\psi^+(X_1,X_2,...)\ Q(X_1,X_2,...) 
\psi'(X_1,X_2,...)$ where $\psi$ refers to the 
proton, $\psi'$ to the antiproton
$X_1$ is the four-point where the first $\bar{q}q$ pair is directly created 
by the photon, $X_2$ the four-point where the second pair is created and 
so on. $Q$ is a kernel describing the $hard$ dynamic steps leading 
from the first 
pair to the other ones (as in PQCD where a gluon radiated from 
a quark splits into another pair). It may only depend on 
the relative positions $x_i$ of the pair creation points. 
Since we have requested that the final state is a non-interacting one, 
soft processes are all factorizable for the time being, that is, they 
take independently place inside $\Phi$ or $\Phi'$. 

%Let us first assume that 
%the virtual photon directly creates a quark-antiquark pair in the four-point 
%$x_1$, as in Fig. \ref{Fig:Feynmann}. 
%Let $\psi'$ and $\Phi'$ 
%refer to the final antiproton, and $\psi^+$ and  
%$\Phi^+$ to the final proton. 
%The four-points $x_2$, $x_3$, and so on, are  
%the positions where other quark-antiquark 
%pairs are created, not directly by the photon. 
%A chain of processes leading from the pair created 
%in $x_1$ to a second pair created in $x_2$ must exist. A 
%standard PQCD example is a gluon radiated from the first quark that 
%generates a second pair, as in Fig.\ref{Fig:particles}. 
%The amplitude for processes like this may 
%be absorbed inside $\Phi'(x_1,x_2,...)$ or $\Phi^+(x_1,x_2,...)$, 
%or appear as a separate function describing the hard part of the 
%process. In this case, other relevant vertex four-points will appear 
%(like the one where a gluon is radiated by a quark line) but they 
%will be integrated over within 
%$\Phi'(x_1,x_2,...)$ or $\Phi^+(x_1,x_2,...)$ separately.  

We may rewrite Eq. (\ref{eq:Amp_charge_start}) for the process $\gamma^*$ 
$\rightarrow$ $\bar{p}p$ 
as:  
\begin{align}
A_{TL,charge}\ =\ &
R_{point,charge}(q,p,\bar{p})\ e_1 \int dX_1 dX_2 .... exp(iqX_1) 
\psi^+(X_1,X_2,...)\ Q(X_1,X_2,...)
\psi'(X_1,X_2,...)\ 
= \label{eq:invariance0}\\
=\ &R_{point,charge}(q,p,\bar{p})\ e_1 
\int dX exp[i(q-p-\bar{p})X] 
\int dx_1 exp(iqx_1) \int dx_2... \delta^4(\sum w_ix_i)\nonumber \\
&\Phi^+(x_1,x_2,...)\ Q(x_1,x_2,...) \Phi'(x_1,x_2,...)\ \equiv
\label{eq:invariance1}
\\
\equiv\ 
&R_{point,charge}(q,p,\bar{p})\ 
\delta^4(q-p-\bar{p})\  
\int d^4x_1 exp(iqx_1) F(x_1), 
\hspace{0.5truecm} x \equiv\ x_1. 
\label{eq:invariance2}
\end{align}

This is simply summarized in Fig. \ref{Fig:Feynmann}, where the virtual photon 
wave is $exp(i q x_1)$ and the blob in $X$ with two legs reaching 
$X_1$ $\equiv$ $X+x_1$ 
is the form factor $F(x_1)$ , that we interpret as 
an amplitude for finding a $\bar{q}q$ pair in $x_1$ resolved level) 
given a $\bar{p}p$ pair in $X$ (unresolved level). 

%%%%%%%%%%%%%%%%%
\section{Soft rescattering}
%%%%%%%%%%%%%%%%%

\begin{figure}
\begin{center}
\includegraphics[width=9cm]{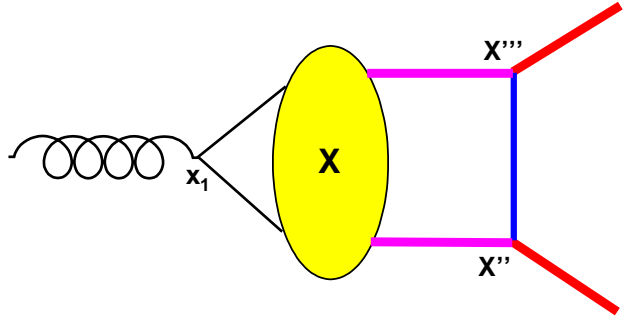}
\caption{
TLFF including t-channel rescattering between the final hadrons. 
} 
\label{Fig:rescattering}
\end{center}
\end{figure}

Let us now consider the t-channel exchange of an interaction quantum 
between the final hadrons, as in 
Fig. \ref{Fig:rescattering}. Rescattering is not necessarily 
elastic, so the hadrons that are produced before rescattering do not 
necessarily coincide with a proton and an antiproton. However, since 
we require the rescattering to be soft, the intermediate state cannot consist 
of, say, three pions or a vector meson. Near threshold, the only cases 
that may be admitted are intermediate states formed by a 
nucleon-antinucleon pair plus possibly one pion. At large $q$ 
we may have near-physical states including a baryon-antibaryon pair. We now 
discuss an example with an intermediate hadron pair.

The initial hadron and antihadron states are created 
in $X$. Let $G(X,X'')$ and $G^*(X,X''')$ be the two-point 
amplitudes describing the propagation of these intermediate-state 
hadrons to the vertexes $X''$ and $X'''$. The amplitude 
$B(X''-X''')$ describes the exchange of an 
interaction between the two. 

Starting from the previous no-rescattering case, we must substitute the  
final state wavefunction $e^{i(p + \bar{p})X}$ with a block 
including the intermediate-state propagations, the interaction, the 
new final state wavefunction:  
\begin{align}
e^{i(p + \bar{p})X}\ \rightarrow 
\int d^4 X'' d^4 X''' 
G(X-X'') G^*(X-X''') B(X''-X''')
e^{-ipX''}
e^{-i\bar{p}X'''} 
\label{eq:rescattering1}
\end{align}
We apply the variable separation:  
\begin{align}
X_R\ \equiv\ (X''+X''')/2\nonumber
\\
\Delta\ \equiv\ X''-X'''\nonumber 
\\
Y\ \equiv\ X_R-X. 
\label{eq:rescattering2}
\end{align}
$Y$ 
is the four-distance between the first hadron-antihadron 
formation point $X$ and the average 
rescattering position $X_R$ $=$ $(X''+X''')/2$. 
So, to further develop Eqs. (\ref{eq:invariance0}-\ref{eq:invariance2}) 
we use 
\begin{align}
e^{iqx_1}\ =\ e^{-iqx} e^{iqX}\ =\ 
e^{-iqx} e^{iq(Y+X_R)},
\label{eq:rescattering3}
\end{align}
\begin{align}
e^{ipX''}\ =\ e^{ip(X+\Delta/2)},\ 
e^{i\bar{p}X'''}\ =\ e^{i\bar{p}(X-\Delta/2)},
\label{eq:rescattering4}
\\ 
G(X''-X)\ =\ G(Y+\Delta/2),\ G^*(X'''-X)\ =\ G^*(Y-\Delta/2),
\label{eq:rescattering5}
\\
B(X''-X''')\ =\ B(\Delta). 
\label{eq:rescattering6}
\end{align}
The 
factor $\int d^4 x e^{iqx}F(x)$ is formally untouched by the previous operations 
(although it may now refer to hadrons different from $p$ and $\bar{p}$) 
and we have 
\ba
\int d^4X d^4 X'' d^4X''' .... \ &=&\\
 \nn
&&\int d^4X_R e^{i(q-p-\bar{p})X_R}\  
\int d^4 Y e^{iqY}\  
\Bigg(\int d^4\Delta G(Y+\Delta/2) G^*(Y-\Delta/2) B(\Delta)\ \Bigg).
\label{eq:rescattering7}
\ea
The 
amplitude that replaces Eq. (\ref{eq:invariance2})
is 
\begin{align}
A(q)\ =\ \delta^4(q-p-\bar{p})\ R_{point\ charge}(q)\ F_0(q)\ F_R(q),
\label{eq:rescattering8}
\end{align}
where 
$F_0(q)$ is the no-rescattering FF corresponding to the creation of the 
hadrons of the intermediate state, and rescattering is described by 
\begin{align}
F_R(q)\ \equiv\ \int d^4 Y\ e^{iqY}\ 
\int d^4\Delta G(Y+\Delta/2) G^*(Y-\Delta/2) B(\Delta).
\label{eq:rescattering9}
\end{align}

Fig. \ref{Fig:rescattering} may describe the two-step process 
$\gamma^*$ $\rightarrow$ $\bar{p}p$ $\rightarrow$ $\bar{p}p$, but 
may also describe a process like 
$\gamma^*$ $\rightarrow$ $\bar{n}n$ $\rightarrow$ $\bar{p}p$ where 
a charge exchange takes place in the second step. This process 
is quite likely near the threshold. At larger $q$, other processes 
with two hadrons in the intermediate state are possible, or with 
two hadrons and some meson. 

Therefore, near threshold we could sum over three diagrams: direct production 
as in Fig. \ref{Fig:Feynmann}: elastic rescattering and $\bar{n}n$ 
intermediate state and we may 
write:
\begin{align}
A(q)\ =\ \delta^4(q-p-\bar{p})\ R_{point\ charge}(q)\ \Bigg(
F_{0,p}(q)\ \Big(
1\ +\ F_{R,diag}(q) \Big)
+\ F_{0,n}(q)\ F_{R,cex}(q),
\ \Bigg)
\label{eq:rescattering12a}
\end{align}
where 
$F_{0,p}(q)$ is the no-rescattering FF for 
$\gamma^*$ $\rightarrow$ $\bar{p}p$, while 
$F_{0,n}(q)$ is the no-rescattering FF for 
$\gamma^*$ $\rightarrow$ $\bar{n}n$. 
$F_{R,diag}(q)$ and $F_{R,cex}(q)$ are the factors describing the 
effects of rescattering without and with charge exchange. 

At larger $q$ more processes will need to be 
included: 
\begin{align}
A(q)\ =\ \delta^4(q-p-\bar{p})\ R_{point\ charge}(q)\ \Bigg(
F_0(q)\ \Big(
1\ +\ F_{R,diag}(q) \Big)
+\ \sum_i F_i(q)\ F_{R,i}(q)
\Bigg),
\label{eq:rescattering12}
\end{align}
where the ``i'' terms take into account inelastic transitions 
proceeding through the intermediate hadron states $|i\big>$ 
formed in the early stages of the process by the amplitude $F_i(q)$. 
We remark that, although these states may be $BB^*$ with large mass 
baryons, here we are only summing over small-virtuality states, that able to 
propagate for a relatively long time. 
%%%%%%%%%%%%%%%%%%%%%%%%%%%%%
\section{Unitarity contribution: a strong imaginary part?} 
\begin{figure}
\begin{center}
\includegraphics[width=9cm]{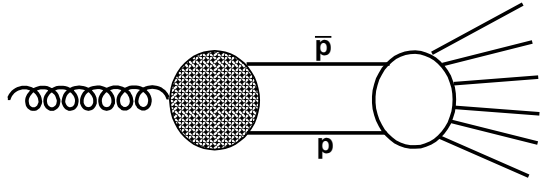}
\caption{
Inelastic diagram of one of the many and frequent processes where the 
just created $\bar{p}p$ pair is lost because of later annihilation into
several pions.} 
\label{Fig:Unitarity}
\end{center}
\end{figure}

%%%%%%%%%%%%%%%%%%%%%%%%%%%%%
Actually the 
most relevant rescattering process is likely to be the one represented in Fig. \ref{Fig:Unitarity}, although it does not lead to a final $\bar{p}p$ state. 
Because of unitarity, this process removes $\bar{p}p$ 
pairs from the outgoing flux of the $\gamma^*$ $\rightarrow$ $\bar{p}p$ 
and from the incoming flux of the $\bar{p}p$ $\rightarrow$ $\gamma^*$ 
processes. So, we expect it to decrease, perhaps heavily, 
the TLFF, but  more subtle and interesting properties 
can be predicted.  

\begin{figure}
\begin{center}
\includegraphics[width=9cm]{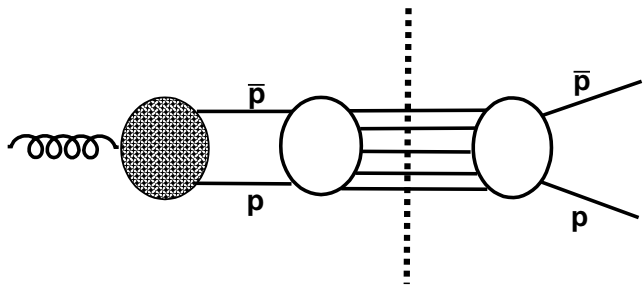}
\caption{
Contribution of the $\bar{p}p$ annihilation processes to the imaginary 
part of the TLFF.} 
\label{Fig:Imaginary}
\end{center}
\end{figure}

In the perturbative Born-style treatment  
of a process at the lowest order including interaction (as for the case 
of previous eq.\ref{eq:rescattering12})  
each accessible final channel just adds itself to the overall cross 
section without disturbing the other channels, because the perturbative 
expansion is based on the requirement of small reaction amplitudes.  
Implicitly this smallness requirement implies 
that the incoming flux is so large that 
when we analyze a process we may neglect the flux depletion due to all 
the other ones. At higher order this inter-channel influence appears as 
an interference effect. 

Evidently a first order perturbative expansion is suitable to small-amplitude 
processes where convergence is fast and higher orders are not decisive. 
When the amplitudes of different processes are so 
large to absorb a relevant portion of the incoming flux,   
they end up competing for this flux, so that each process reduces the 
cross section of the other ones. This is well visible in all those 
hadronic processes where we find a cross section minimum at an energy $E_0$ 
because of the competition 
with a channel that is resonant at $E_0$. 

The inclusive cross section of the $\bar{p}p$ annihilation  
at relative momenta $\sim$ 10-1000 MeV/c is a relevant portion of the 
unitarity limit (that takes place when all the incoming 
flux is completely absorbed in a partial wavea). 
Since inelastic processes like those 
of Fig. \ref{Fig:Unitarity} present a large cross section, they 
also give a large contribution to the imaginary part of the elastic 
$\bar{p}p$ $\rightarrow$ $\bar{p}p$ scattering, and so, indirectly,  
to the imaginary part of the $\gamma^*$ $\rightarrow$ $\bar{p}p$, 
due to the Cutkowsky cut that can be seen in 
Fig. \ref{Fig:Imaginary}. 

To clarify this point, if $\bar{p}p$ annihilation saturated 
the unitarity limit, then the $\bar{p}p$ scattering amplitude 
would be completely imaginary (it would exactly coincide with the 
center of the Argand circle). Watson's theorem \cite{Watson:1954uc} 
%nuova (K.M.Watson, Phys.Rev.{95}, 228 (1954) )
states that  
the phase of an amplitude leading to a final
state with two strongly interacting particles in a given 
partial wave is
the same as the scattering phase of that pair. 
So, near the threshold (pure S-wave) 
even the overall $\gamma^*$ $\rightarrow$ $\bar{p}p$ 
amplitude would be imaginary. 

Writing the amplitude for 
$\gamma^*$ $\rightarrow$ $\bar{p}p$ as the amplitude for a ``pointlike'' 
process like $\gamma^*$ $\rightarrow$ $\mu^+\mu^-$ times the TLFF 
of the proton, 
and observing that the pointlike amplitude presents no special features, the 
imaginary character of the overall amplitude would entirely derive 
from an imaginary TLFF. 

%We cannot re-use 100 \% of what is known from the theory and the 
We cannot implement all what is known from the theory and the 
experiment concerning $\bar{p}p$ scattering and annihilation, because 
the TLFF only involves 
the vector component of the $\bar{p}p$ flux. However, the fact that the 
cross section for many-pion 
annihilation is large should be a common property of all the 
$\bar{p}p$ partial waves. So, we can consider some relevant results 
from the literature of $\bar{p}p$ interactions at small energies, associated 
with this property. 

First, we know from theory and have experimental check from antiprotonic 
atoms that the real and the imaginary part of the scattering amplitude 
near the threshold are equal (this equality takes place in regime 
of S-wave dominance, that is between $p$ $=$ 0 and $p$ $\sim$ a few 
tenths of MeV/c corresponding to a relative energy $\approx$ 1 MeV). 
The inelastic cross section is large, but still far from 
the unitarity limit. 

At $p \gtrsim 200$ MeV/c (relative energy $\gtrsim$ 40 MeV) 
an approximately eikonal regime 
is present, where the physics is essentially ``black-sphere plus 
diffraction''. In this case $\bar{p}p$ scattering is mainly 
forward and imaginary-dominated, and the radius of the black sphere is  
of magnitude $simeq$ 1 fm and scarcely energy dependent. No resonances have 
ever been seen. 

We can draw some hypotheses from this. 
In the eikonal regime ($p \gtrsim 200$ MeV/c) rescattering 
is relevant but cannot dominate the problem, because its geometrical 
cross section $<$ 100 mbarn $=$ 10 fm$^2$ (it tends to 50 mbarn 
at GeV energies) limits the probability 
that a proton and an antiproton rescatter, considering that 
we assume that  ``rescattering''  takes place when the hadrons are distant at least 1 fm. 
On the other side, these numbers 
also mean that rescattering is present and not 
negligible at all. 
This black-disk rescattering implies  
regeneration of flux by diffraction of the same magnitude. 
So it does not decrease the TLFF, but rather shifts its phase. 
The regenerated flux presents 
a phase shift of $90^\circ$ with respect to the non-scattered flux. So, 
it introduces a relevant, although not dominant, imaginary part in 
the TLFF. 

Near threshold, when the S-wave is relevant or dominant, one can apply 
Watson's theorem and, as already observed, the TLFF must present the 
same phase features of the $\bar{p}p$ scattering amplitude, $i.e.$, equal 
real and imaginary parts.

%%%%%%%%%%%%%%%%%%%%%%%%%%%%%%%%%%%%%%%%%%%
\section{Time and distance in three-dimensional and non-relativistic sense}
%%%%%%%%%%%%%%%%%%%%%%%%%%%%%%%%%%%%%%%%%%%

It is very unlikely that the ``bare formation'' process (involving 
ultrarelativistic quarks and gluons) 
and soft hadron-hadron rescattering 
are theoretically handled in a similar way near the threshold, 
where the rescattering 
problem is  non-  or a weakly relativistic. For treating 
proton-antiproton rescattering and calculating the correcting factors 
$F_{R,i}$, a more suitable variable is $r$, 1/2 of the proton-antiproton 
distance, 
to be used in a nuclear physics style formalism, like e.g. Glauber's. 

Apparently $r$ has no connection with time evolution. However, 
a strong correlation is likely to exist between $r$ and the time interval from 
the beginning of the hadron formation 
when both quantities are soft-scaled and in a kinematical regime not too close to the 
threshold. Although there is a quantum correlation 
between $r$ and $t$, this correlation is classical  
($r$ $\approx$ $vt$) within an uncertainty $\delta r$ $\sim$ $1/p$, 
where $p$ is the relative three-momentum of the pair. For $p>$ 200 MeV/c (corresponding to $\delta r \approx$ 1 fm) the 
time-distance correlation is classical within 1 fm, 
the hadron radius. So, at a given time, the hadron-hadron separation assumes  
a reasonably well defined value. Evidently this argument 
is not valid for $p \ll $ 200 MeV/c or for $t \ll $ 1 fm$/v$, but makes 
sense within these exceptions. 

From now on, let us distinguish between $R$, the true quantum 
variable indicating 
1/2 of the $\bar{p}p$ distance, and $r$, that we define as 
\ba
&r\ &=\ vt,\\
&q\ &=\ 2 E_p\ =\ 2M/\sqrt{1-v^2},\\
&v\ &=\ \sqrt{1 - {{4M^2} \over q^2}}.
\ea
($v$ is here meant as $v/c$, and $t$ as $ct$). 

It
is important not to forget in the following that, because of the 
deterministic relation $r=vt$, $r$ is just an alias for $t$ 
in the Fourier transforms, 
although the quantum variable $R$ may satisfy 
$< R >$ $\approx$ $vt$ within the above discussed limits. 

One can rewrite the time Fourier 
transform (introduced above to define the FF  
in terms of spacetime functions) in terms of $r$.

In the term $e^{iqt}$ we have 
\be
qt\ =\ qr/v\ =\ 
{q \over {\sqrt{1-4M^2/q^2}}} r. 
\label{eq:nonrelativistic1}
\ee
The Fourier transform may be performed with respect to 
\be
q'\ \equiv\ q - 2M,
\ee
instead of $q$. 
\be
(q-2M)t\ =\ q 
{{q-2M} \over {\sqrt{q^2-4M^2}}} r\ =\sqrt{ {q-2M} \over {q+2M} }\ q r.
\label{eq:nonrelativistic2}
\ee

In the Fourier transform, when $t$ is substituted by 
$r$, all the functions of $t$ need to be expressed as functions of $r$. 
Now, $r$ can only be positive. We notice  
that in spite of appearances even $t$ can have one sign only, in 
each independent process $TL_\pm$ 
described by $F(t)$ ($\bar{p}p$ creation or annihilation,  
see Eqs. (\ref{eq:eq1},\ref{eq:eq2}).
Since $t$ is the time of the 
initial quark pair creation or of the annihilation of the final 
pair, on a time 
axis where the origin is the creation or annihilation time for 
the whole $\bar{p}p$ system, 
positive $t$ describe $\bar{p}p$ annihilations where ``rescattering'' 
means initial state interactions, while negative $t$ the 
reverse process where ``rescattering'' means final state interactions. 

Let $\vec P$ be the three-momentum of the 3-dim Fourier transform: 
\begin{align}
\int d^3\vec r F(r) e^{i\vec P \vec r}\ 
=\ \frac{2\pi}{iP}\ 
\int_0^\infty\ \Big(e^{i P r}\ - e^{-i P r}\Big)\ F(r)\ rdr . 
\label{eq:rt2prel}\\
\end{align}
In 
this obvious sequence of passages it is implicitly assumed that 
$F(r)$ does not 
depend on the reciprocal orientation of $\vec r$ and $\vec P$. It is however 
evident that diverging and converging waves correspond to the channels 
$TL_\pm$ of Eqs. (\ref{eq:eq1},\ref{eq:eq2}). 
Indeed, $F(r)$ describes 
di-hadron state configurations, not di-lepton ones. Only in 
the elastic process $\bar{p}p$ $\rightarrow$ $\bar{p}p$ we have the 
simultaneous presence of converging and diverging di-hadron waves. 

Let us consider the $possibility$ of a 
difference between the FFs associated to 
converging and diverging waves $i.e.$, to the processes 
of hadron pair annihilation and 
creation. 
\ba
F(P)\ \equiv\ &
\int d^3\vec r F(r) e^{i\vec P \vec r}\ 
=\ \frac{2\pi}{iP}\ 
\int_0^\infty\ \Big(e^{i P r}\ - e^{-i P r}\Big)\ F(r)\ rdr \\=\  
&\frac{2\pi}{iP}\ \int_{-\infty}^\infty\ \Big[\theta(r)F(r) + \theta(-r)F(-r)\Big] 
e^{i P r}\ rdr \label{eq:previous}
\ea
Writing the initial integral $F(P)$ in terms of $F(r)$ and $F(-r)$ 
is just formal, since we have substituted $r$ with $-r$ in the second 
integral where $-r$ is anyway positive because of $\theta(-r)$. 

Now we introduce by assumption a change: we modify (\ref{eq:previous}), assuming 
that converging and diverging $\vec P$ may correspond to different 
FFs $F_\pm(r)$: 
\ba
Eq. \ (\ref{eq:previous})\ 
\longrightarrow\ 
\frac{2\pi}{iP}\ \int_{-\infty}^\infty\ 
\Big[\theta(r)F_+(r) + \theta(-r)F_-(r)\Big] 
e^{i P r}\ rdr 
\label{eq:later}
\\
=\ \frac{2\pi}{iP}\ \int_{-\infty}^\infty\ \Big\{
F(|r|)\ 
+\ \frac{1}{2}\Delta(|r|)\big[\theta(r)\ -\ \theta(-r)\big] 
\Big\}
e^{i P r}\ rdr \\
\equiv\ F_{sym}(P)\ +\ F_{asym}(P), \ 
\Delta(r)\ \equiv\ F_+(r)-F_-(r).
\label{eq:rt2}
\ea
The two integrals present a 90$^\circ$ relative phase because the former 
is even while the latter is odd in $r$. So if for example the former is real,  
the latter must be imaginary. 
When $\Delta(r)=0$, $F(P)$ coincides with $F_{sym}(P)$. 

Written this way, the Fourier transform is one-dimensional and its 
integration runs from $r=-\infty$ 
to $r=+\infty$, not to become a Laplace transform with 
limited convergency properties. It better reproduces the 
correspondence $r\leftrightarrow t$ that was the starting point 
of this section. 

To draw a clearer borderline between 
form and substance, we remark that ``negative $r$'' only means  
``negative $Pr$'' in $exp(iPr)$. 
$P$ is defined in such a way to have, according to the previous 
equations,  
\begin{align}
Pr\ =\ (q-2M)t 
\end{align}
In our 
Fourier transform prescription, $q$ means $q_0$ (the zeroth component 
of $q^\mu$ in the c.m. frame) 
and is negative in one of the two reactions $TL_\pm$, 
see Eqs. (\ref{eq:table2}) and (\ref{eq:table3}). 
Correspondingly, $P$ is negative in one of the two reactions, 
but here the negative sign has been 
transferred to $r$, leaving $P$ positive defined in both reactions. 
The relevant 
sign is the one of $Pr$, and the physical properties of the system allow 
to calculate the 
Fourier integral from $Pr$ $=$ $-\infty$ to $Pr$ $=$ $+\infty$. Our 
``negative $r$ prescription'' only makes this range property explicit. 

\ba
&Final\ \bar{pp}:&\nonumber\\
&reality:&\ \ Pr\ >\ 0,\  P > 0,\ r > 0;\\
&formal:&\ as\ in\ reality.\\ 
&Initial\ \bar{pp}:&\nonumber\\
&reality:&\ \ Pr\ <\ 0,\  P < 0,\ r > 0;\\
&formal:&\ \ Pr\ <\ 0,\  P > 0,\ r < 0.
\ea

It is evident that using the variables $P$ and $r$ is not optimal  
to describe the full relativistic development of the FF. However it 
is practical to recover Ref. \cite{Bianconi:2016bss}, where 
the rescattering FF, $F_R$, in Eq. (\ref{eq:rescattering8}) is calculated 
as a function of $P$ using a 
traditional $r-$dependent non-relativistic format for 
low-energy ${\bar{p}p}$,   
while the leading term $F_0(q)$ may be taken as a function of $q$  
from one of the above quoted parameterizations or models. 

%In eq.\ref{eq:rescattering9} 
%$Y$ is the time separating rescattering from the leading quark-level 
%process. The softness requirement means   
%that rescattering takes place on a relatively long time scale. Given 
%this condition, it looks appropriate substituting $exp(iqY)$ with 
%$exp(iPr)$ and developing $F_R(q)$ in terms of $P$ and $r$. 
%%%%%%%%%%%%%%%%%%%%%%%%%%%%%%%%%%%
\section{Optical potentials and inclusive absorption in rescattering.}
%%%%%%%%%%%%%%%%%%%%%%%%%%%%%%%%%%%
%Both the charge distribution $\rho(\vec x)$ and the ``time charge'' 
%distribution $R(t)$ derive from integrals of an underlying fourier 
%transform $F(x)$. We know that $\rho(\vec x)$ must be real, while 
%$R(t)$ has not such a constraint. For exapmle, we could have 
%a complex $F(x)$ whose imaginary part is odd in time, but not in space. 
%Then $\rho(\vec x)$ $=$ $\int F(x) dt$ is real, while 
%$R(t)$ $=$ $\int F(x) d^3x$ may be complex.  

%Alternatively in the previous section we have seen that,  
%starting from an $F(x)$ that is real and not complex, 
%an imaginary part $F_{asym}(P)$ in Eq. (\ref{eq:rt2}) is obtained as 
%a consequence of a term $\Delta(r)= F_+(r)-F_-(r)$ expressing a time 
%reversal asymmetry (diverging and converging hadron waves are 
%not associated to the same FF). This introduces a time reversal asymmetry. 
In Eq. (\ref{eq:later}) we have introduced a possible difference in the 
form factors associated to annihilation or creation of a $\bar{p}p$ 
pair from/into a photon. 

Time reversal invariance forbids any asymmetry in an amplitude involving 
really pure initial and final states. 
We need to remind that the FF is not the amplitude of a measurable process, 
but rather a hadron-to-constituent splitting amplitude. In spacetime 
representation, given a proton 
and an antiproton overlapping in the origin, $F(x)$  
is the amplitude for 
having a quark-antiquark pair in $x$. 
This corresponds to a process that is not 
observable. What is observable is a process of which the 
$\bar{p}p$ $\rightarrow$ $\bar{q}q$ splitting is a subprocess. 

If a time asymmetry is present, this must 
be related to an incomplete 
control on the purity of the states involved in this splitting. Here 
we give an example of how this is produced in rescattering, 
using a very basic optical potential modeling of rescattering. 

We consider space-homogeneous rescattering consisting 
of inclusive annihilations, 
like for antiprotons in nuclear matter. This corresponds to a 
time- and position-independent optical potential 
dominated by its imaginary part. We neglect the real part. 
Actually $\bar{p}p$ annihilations do not take place 
everywhere with the same probability, but only when the two 
hadrons are within a few fm, so we will later correct for this. 
What simplifies the homogeneous potential treatment is that we 
already know the solution of the problem. 

In Eqs. (\ref{eq:previous},\ref{eq:rt2}), 
$exp(iPr)$ is the wavefunction of the relative motion of the 
hadron pair, in absence of rescattering. Let us correct this by a 
damping factor
\begin{align}
e^{- K't}\ \equiv\ e^{-Kr}\ =\ e^{\pm K|r|},
\end{align} 
distorting this wave function, where 
%The coefficient $K'$ and $K$ 
%can be easily related each to the other via eqs. \ref{eq:nonrelativistic1}
%and \ref{eq:nonrelativistic2}
\begin{align}
K\ =\ 1/\lambda,\ 
K'\ =\ v/\lambda, 
\end{align}
$\lambda$ is the inclusive annihilation free mean path, and $v$ the 
relative velocity. 
So the full wave is 
\ba
exp(iPr\pm K|r|)
\ea
where the sign $+$ is for negative $r$. 
We notice that $exp(-K't)=exp(-Kr)$ 
is not time symmetric. Whether hadrons are converging or diverging, 
at later times their flux is smaller. 

Without some boundary condition, the above exponentials lead to convergence 
problems at large $|r|$. 
So we now modify the approximation, to take into account the finite 
size of the absorption region and to focus on the leading 
even and odd terms: 
\begin{align}
exp(- Kr)\ \rightarrow\ 1 - Kr\ for\ |Kr|\ <\ 1, \nonumber \\
exp(-Kr)\ \rightarrow\ 2\ for\ Kr\ <\ -1, \nonumber \\ 
exp(-Kr)\ \rightarrow\ 0\ for\ Kr\ >\ 1. 
\label{eq:damping}
\end{align}
This is equivalent to assume that the absorption region is of 
size $1/K$ $=$ $\lambda$. When Eq. (\ref{eq:damping}) 
is inserted into the Fourier transform (\ref{eq:rt2}), we obtain two 
integrals of functions of opposite $r-$parity. This fits the 
general scheme of Eq. (\ref{eq:later}) and we obtain two pieces 
with a relative phase of $90^\circ$. 

We have normalized the incoming and outgoing hadron waves to be 
the same in the origin, that is proper in a scattering-from-a-center 
problem. We could improve our treatment by assigning two different 
normalizations for the incoming and outgoing waves (since we 
need handling two different reactions, not two different stages 
of the same one). This 
would not change our point: the distorting factor would not be 
$t-$symmetric. Incoming waves would have larger flux 
at larger $r$, outgoing waves at smaller $r$. 
As a consequence the final result would 
contain two pieces whose relative phase is $90^\circ$ in any case. 

A realistic optical potential changes a plane wave $exp(iPr)$ in 
a way that is 
qualitatively similar to the above considered case. The result of 
its action is exactly $exp(iPr-Kr)$ if the potential is spatially 
homogeneous with the form $V$ $=$ $-i\hbar K^2/2m$. In a more general 
case the wave function is more complicate but as a rule  
it does not respect unitarity: the particle flux decreases.   
For the flux density vector $\vec J$ we have 
$\partial \rho / \partial t + div(\vec J)$ $<$ 0. 

From a formal point of view, 
being complex, the optical potential makes the hamiltonian not hermitian. 
So a basic prerequisite for both unitarity and time reversal 
symmetry is not present. It is the interaction itself that 
becomes intrinsically not T-invariant, due to how the potential 
is built from interactions that  are of course T-invariant. 

The physical reason underlying the loss of hermiticity 
is that the lost flux is related with the 
transformation of the 
$\bar{p}p$ initial state into something else, but this ``something else'' 
includes too many possibilities for a 
coherent reversal to be possible within a finite time. 

The optical potential is used in an equation (for example, Schroedinger 
equation) whose solution is the wave function relative to a single channel, 
in our case $\bar{p}p$.  
The physical reality behind this is a coupled set of equations 
where all the channels 
that may be the result of an interaction between $\bar{p}$ and $p$ should  
be included, for example $\bar{n}n$ or several 
pions. The most simple problem of this family is 
two coupled one-dimensional channels. Here 
we have periodic oscillation of the flux between 
the two channels. 
Although it is formally stated in the Fock-Krilov theorem \cite{Fock:1974em}, 
%Fok-Krilov theorem (cite-xxxx-FK) 5V. A. Fock and S. N. Krylov, 
%Zh. Eksp. Teor. Fiz. 17, 93 (1947).
intuitively the time 
evolution of a system is periodic or quasi-periodic if the starting 
channel couples to 
a discrete and finite set of channels. If these channels form a continuum, 
starting at $t = 0$ from a configuration 
where all the flux is in one channel, the later evolution is 
irreversible and depletes this channel. 

Let us consider a common final state of $\bar{p}p$ annihilation, that is 
six pions. The amplitude for 
$\bar{p}p \rightarrow 3 \pi^+3\pi^-$ is T-invariant, but the phase 
space enormously advantages the direct reaction over the reverse one. 
Any $effective$ amplitude elastically leading the $\bar{p}p$ channel into 
itself must take this phase space unbalance into account, but this means 
to convert a probabilistic effect into an amplitude. So this amplitude 
is not a proper one. 

From another point of view, bypassing the optical potential treatment, 
one can obtain the damping factor $exp(-K't)$ by assuming 
that 

(i) rescattering implies a probability $2K'dt$ of 
losing the $\bar{p}p$ pair within the short time range $dt$ because of 
an annihilation event, 

(ii) two or more annihilation 
events involving the same flux of colliding particles are incoherent 
(classical physics, where ``flux'' means many 
unrelated $\bar{p}p$ pairs). 

If we sum over 0, 1, 2, ... $\infty$ rescattering events, 
Poisson's law gives $exp(-2K't)$ as 
a probability of zero annihilation events. So $exp(-2K't)$ is  
also the $\bar{p}p$ survival probability after a  
time $t$. Then its square root, $exp(-K't)$, may be interpreted as 
an effective survival amplitude. 

We notice that although $exp(-K't)$ has been calculated by a probabilistic 
method, it is a quite reasonable form for the amplitude of remaining in the 
initial channel. This ``reasonable'' character derives from the 
fact that annihilation may take place is so many different ways that we 
can apply random phase arguments to all those processes where  
$\bar{p}p$ pairs are restored in events like e.g. $\bar{p}p$ 
$\rightarrow$ $\bar{n}n$ $\rightarrow$ $\bar{p}p$. If annihilation could 
only take place into e.g. $\bar{n}n$, the factor $exp(-K't)$ would 
be wrong. 

We observe that a basis of our previous work \cite{Bianconi:2016owb} was 
the idea that this no-correlation assumption was not 100 \% true, and 
a short-distance coherent coupling with some channels might lead to 
observable sub-leading effects. However, we expect that 
the leading effect of rescattering is a loss of $\bar{p}p$ 
pairs. 

\section{Detection of 
differences between the proton TLFF in the creation and annihilation 
processes}

Summarizing the previous discussion, inelastic rescattering may 
create a difference between 
the proton effective TLFF as seen in the $\bar{p}p$ creation or annihilation 
channels. Being this difference related to rescattering, we don't expect 
it to be large, but it may exist. We do not expect it to be observable 
in a straightforward way. 

We must remind that the TLFF alone 
is the amplitude for an unobservable process (a hadron-to-quark  
splitting), that must be convoluted with the 
amplitudes for the other parts of an observable process. The overall 
process must respect time reversal symmetry in the form of detailed 
balance (the cross sections of the reactions $e^+ + e^- \to \bar p +p $ 
and $\bar p + p \to e^+ + e^-$ must be equal once phase space factors 
have been removed), so indirect constraints on the TLFF are present. 

In a cross section the effective TLFF appears multiplied by its 
complex conjugate, so the only left part of the process amplitude 
is the pointlike process amplitude: $|A_{full}|^2$ $\propto$ 
$|F|^2 |A_{point}|^2$, see Eq. (\ref{eq:Amp_start}). 
Since both, the full and the pointlike processes, are time-reversal 
symmetric, no difference in the ``creation'' and ``annihilation''  
TLFF may be detected in this kind of 
observable. This suggests that $|F|^2$ must be the same for the two 
cases, and a difference may only be present in the phases. 
A phase difference may be seen in interference observables where 
an individual FF appears linearly. Given the relevance of phases, 
such effects should be searched for in the individual electric and 
magnetic form factors, and not in the effective one, whose phase is 
not well defined. Feasible measurements of interference effects in 
TLFF,  were first suggested in Ref. \cite{Dubnickova:1992ii} for the reaction (\ref{eq:eq1a}), 
and in in Refs. \cite{Bilenky:1993cd,TomasiGustafsson:2005kc} for the reaction (\ref{eq:eq2a}).
Particularly interesting and sensitive to nucleon models are single spin polarization effects.

\section{Conclusions}

This work discusses the role of soft rescattering in the calculation of 
small corrections to the proton timelike FF. The present analysis has been carried on within the spacetime 
scheme developed by us in a series of previous works. 

The corrections to the FF are of two kinds: 

a) Rescattering leading to a final $\bar{p}p$ state, 
anticipated by an intermediate stage where a $\bar{p}p$ state or a 
different hadron state is present. The corrected form of the TLFF 
is simple in principle, but it is important 
to understand how many intermediate states may play a role. Near 
threshold they are restricted to nucleon-antinucleon states. 

b) Unitarity corrections, in particular annihilation of the $\bar{p}$ 
and $p$ into a multi-meson state. The phenomenology of $\bar{p}p$ 
scattering suggests that this is the most relevant phenomenon 
associated to rescattering. It should contribute to produce a relevant 
imaginary part of the TLFF, that  could be as large 
as the real part near the reaction threshold. 

Since it is very likely that the theoretical treatment of $\bar{p}p$ 
rescattering follows traditional nuclear physics lines, we have 
matched the variables that appear in this case with the standard 
variables appearing in a relativistic treatment of the TLFF. 

We have shown that some constraints on TLFF  as time invariance 
are weakened by the fact that the FF itself is the amplitude 
of a process that cannot be observed ``stand-alone''. This 
creates the interesting possibility that the TLFF appearing in 
creation and annihilation of the $\bar{p}p$ pair may differ. 
We have shown that a difference would customarily appear in 
a standard optical potential treatment of rescattering. Because 
of symmetry constraints on the full process, this difference is 
expected to concern phases and be visible in interference processes. 

%\begin{thebibliography}{35}

%\end{thebibliography}
%\input{rescattering.bbl}
\bibliography{Biblio}
\end{document}